\documentclass[]{PoS}
\usepackage[utf8]{inputenc}
\usepackage{amsmath}
\usepackage{mathtools}
\usepackage{slashed}
\usepackage{bbm}
\usepackage{mleftright}
\usepackage{nicefrac}
\usepackage[
	style=numeric-comp,
	backend = bibtex,
	bibencoding = utf8,
	sorting=none,
	firstinits=true,
	isbn=false,
	url=false,
	doi=false,
	maxbibnames = 99
	]{biblatex}
\DeclareFieldFormat
  [article,inbook,incollection,inproceedings,patent,thesis,unpublished,suppbook,suppcollection,suppperiodical]
  {title}{\mkbibemph{#1}}
\DeclareFieldFormat{eprint:arxiv}{[#1]}

\renewbibmacro{in:}{}

\newcommand{\dx}{\mathrm{d}}
\newcommand{\Dx}{\mathcal{D}}
\newcommand{\ii}{\mathrm{i}}
\newcommand{\Nf}{{N_\mathrm{f}}}
\newcommand{\Nfc}{{N_\mathrm{f}^\mathrm{cr}}}

\newcommand{\ex}[1]{\mathrm{e}^{#1}}
\newcommand{\srm}[1]{_\mathrm{#1}}

\title{Critical flavour number of the Thirring model in three dimensions}

\ShortTitle{Critical flavour number of the Thirring model in three dimensions}

\author{\speaker{Daniel Schmidt}\\%
      Theoretisch-Physikalisches Institut, Friedrich-Schiller-Universit\"at Jena, 07743 Jena, Germany\\
        E-mail: \email{d.schmidt@uni-jena.de}}

\author{Björn Wellegehausen\\
       Institut f\"ur Theoretische Physik, Justus-Liebig-Universit\"at Giessen, 35392 Giessen, Germany and\\
      Theoretisch-Physikalisches Institut, Friedrich-Schiller-Universit\"at Jena, 07743 Jena, Germany
       E-mail: \email{bjoern.wellegehausen@uni-jena.de}}

\author{Andreas Wipf\\
       Theoretisch-Physikalisches Institut, Friedrich-Schiller-Universit\"at Jena, 07743 Jena, Germany\\
       E-mail: \email{wipf@tpi.uni-jena.de}}       
       
\abstract{
The Thirring model is a four fermion theory with vector interaction.
We study it in three dimensions, where it is closely related to QED and other models used to describe properties of graphene.
In addition it is a good toy model to study chiral symmetry breaking, since a phase with broken chiral symmetry is present for the model with one fermion flavour.
On the other hand, there is no such phase in the limit of infinitely many fermion flavours.
Thus, a transition at some critical flavour number $\Nfc$ is expected, where the broken phase vanishes.

The model was already studied with different methods, including Schwinger-Dyson, functional renormalization group and lattice approaches.
Most studies agree that there is indeed a phase transition from a chirally symmetric phase to a spontaneously broken phase for a small number of fermion flavours.
But there is no agreement on the critical flavour number and further details of the critical behaviour.
Values of $\Nfc$ found in the literature usually range between 2 and 7.

All earlier lattice studies were performed with staggered fermions, where it is questionable if the continuum limit of the lattice model has the same chiral symmetry as the continuum model.
We present an approach for simulations of the Thirring model with SLAC fermions.
With this choice, we can be sure to implement the full chiral symmetry of the continuum model.
First results from simulations are shown but do not allow a reliable estimate of $\Nfc$ so far.
}

\FullConference{The 33rd International Symposium on Lattice Field Theory\\
		 14 -18 July  2015\\
		 Kobe International Conference Center, Kobe, Japan}

\begin{document}
\section{Introduction}
\label{s:introduction}
The Thirring model was introduced in 1958 by Walter Thirring as a two-dimensional quantum field theory, when he was looking for an exactly solvable theory with interacting fermions \cite{Thirring1958}.
Its main feature is a four fermion interaction with a current term.
While its original formulation in two dimensions is quite well known \cite{Sachs1996}, there are a lot of open questions in a three-dimensional spacetime with a variable number of $\Nf$ fermion flavours. 
The Lagrangian in Euclidean spacetime is given by
\begin{equation}
  \mathcal{L}=\bar\psi_j \ii \slashed{\partial} \psi_j-\frac{g^2}{2\Nf}\left(\bar\psi_j \gamma^\mu \psi_j\right)^2 \qquad j=1,\dots,\Nf.\label{e:th_orig_lagrangian}
\end{equation}
We employ the usual Hubbard-Stratonovich transformation to introduce an auxiliary field $V_\mu$ and obtain the Lagrangian
\begin{equation}
  \mathcal{L}=\bar\psi_j\ii\gamma^\mu \left(\partial_\mu-\ii V_\mu\right) \psi_j + \lambda_\mathrm{Th} V^\mu V_\mu
\end{equation}
with $\lambda_\mathrm{Th}=\nicefrac{\Nf}{2g^2}$.
We use a four-dimensional representation of the Clifford algebra, which is reducible in a three-dimensional spacetime.
This theory was shown to be renormalizable in a large $\Nf$ expansion for $2<d<4$ \cite{Parisi1975,Hands1995}.
The 3D formulation is useful to study chiral symmetry breaking and is motivated by a strong similarity to QED$\srm{3}$ \cite{Hands1995,Itoh1995}.
These models can be found in the literature to describe the electronic properties of graphene \cite{Semenoff1984,Hands2008} or high temperature superconductors \cite{Franz2002,Herbut2002}.

The most interesting feature of the Thirring model is its chiral symmetry.
The action is invariant under transformations generated by the matrices $\left\{\mathbbm{1}, \gamma_4, \gamma_5, \ii\gamma_4\gamma_5\right\}$.
In addition, there is a flavour symmetry, which leads to a full symmetry group of $U(\Nf,\Nf)$ \cite{Gomes1991}.
This symmetry can be spontaneously broken to
\begin{equation}
  U(\Nf,\Nf)\rightarrow U(\Nf)\otimes U(\Nf). \label{e:th_breaking_pattern}
\end{equation}
The large $\Nf$ expansion shows, that no symmetry breaking occurs for $\Nf\rightarrow\infty$.
On the other hand, there is a correspondence due to a Fierz identity of the \emph{irreducible} Thirring model with $\Nf=1$ to a Gross-Neveu model.
The latter is known to always show chiral symmetry breaking, see for example \cite{Reisz1998}.
Since $\Nf=1$ in the irreducible representation can be thought of as ``$\Nf=0.5$'' in the reducible representation, this can be taken as a hint, that our model shows chiral symmetry breaking for small $\Nf$.
Thus, there must exist a critical flavour number $\Nfc$, such that chiral symmetry breaking is possible for $\Nf<\Nfc$ but not for $\Nf>\Nfc$.
The main motivation of our work is to determine the value of $\Nfc$ from lattice simulations.

There have been numerous other approaches to determine $\Nfc$, which found a variety of values:
The earliest works used Schwinger-Dyson equations and found values of $\Nfc=3.24$ \cite{Gomes1991}, $\Nfc=4.32$ \cite{Itoh1995} or even $\Nfc=\infty$ \cite{Hong1994}\footnote{Although the model is considered in the irreducible representation in this paper.}. 
There is a paper \cite{Kondo1995} constructing an effective potential from the large $\Nf$ expansion, which found $\Nf=2$ and an extensive study \cite{Janssen2012-2} of four fermion theories with methods of the functional renormalization group, where $\Nfc\approx 5.1$ is found.
Lattice studies were performed with staggered fermions only.
Earlier simulations \cite{Kim1996,DelDebbio1997-1,DelDebbio1999} used a HMC algorithm, allowing integer values of staggered fermion flavours, where each lattice flavour corresponds to \emph{two} flavours of the continuum model.
These studies report results for simulations with $\Nf=2,4,6$.
They used a non-vanishing mass, which breaks the chiral symmetry explicitly, and obtained fits to the equation of state, that relates the chiral condensate to the mass.
The authors conclude, that there must be a change in the chiral behaviour between $\Nf=4$ and $\Nf=6$.
Moreover simulations with a Hybrid Molecular Dynamics algorithm were performed with non-integer values of $\Nf$.
The first study \cite{Hands1999} agrees well with the other lattice results, while a more recent investigation \cite{Christofi2007} found $\Nfc\approx 6.6$.
As pointed out in \cite{DelDebbio1999}, there is a difficulty in using staggered fermions: 
The staggered lattice model shows a chiral symmetry breaking pattern different from \eqref{e:th_breaking_pattern} and it is not clear, if the continuum limit of the staggered lattice model leads to the correct breaking pattern.

Another point that may be important to correctly interpret our results, is made in \cite{DelDebbio1997-1, Christofi2007}.
It is shown there, that the renormalizability of the Thirring model in large $\Nf$ expansion requires the vacuum polarization tensor to be transversal (as in QED).
This can be achieved by a coupling constant renormalization with renormalized coupling
\begin{equation}
 g_\mathrm{R}^2=\frac{g^2}{1-g^2 J(m)}, \label{e:renorm_coupling}
\end{equation}
where $J(m)$ is an integral depending on the bare mass $m$ with a value of $\frac{2}{3}$ for $m\rightarrow 0$ in first order of the large $\Nf$ expansion.
Remarkably, the renormalized coupling can get negative, if the inverse of the bare coupling squared gets smaller than $J(m)$.
Then the model is believed to be in an unphysical phase where it is not unitary.
The authors of \cite{Christofi2007} interpret a decrease in the chiral condensate for increasing bare coupling as a transition to this unphysical phase and assume that the position of the maximum corresponds to $g\srm{R}^2\rightarrow\infty$ in equation \eqref{e:renorm_coupling}.
This interpretation is criticized in \cite{Janssen2012-2}, since in the functional renormalization group approach, the transversality of the vacuum polarization is not necessary to ensure renormalizability.

\section{Current Approach}
\label{s:current_approach}
In order to overcome the disadvantages of the staggered fermions, we use the SLAC fermion formulation \cite{Drell1976}.
This is possible here, since the Thirring model is not a gauge theory and the argument considering the non-covariance of the vacuum polarization of \cite{Karsten1979} does not apply here.
It was shown \cite{Bergner2008} that Wess-Zumino models with SLAC derivative are renormalizable and have the correct continuum limit.
In recent years, the nonlocal SLAC fermions have been successfully used in many simulations of Yukawa-type models, see \cite{Wozar2011} and references therein.
We take this as a strong indication that the SLAC derivative works here, too.
In position space, it is given by
\begin{equation}
  \partial^{\mathrm{SLAC}}_{xy}=
  \begin{cases}
    0			& x=y\\
    \frac{\pi}{L}(-1)^{x-y}\frac{1}{\sin\mleft(\frac{\pi}{L}(x-y)\mright)}		& x\neq y                                                                                                                                              
  \end{cases}
\end{equation}
while the application in momentum space is per definition just a multiplication with the lattice momenta.
There are no doublers and we have the exact chiral symmetry.

We can perform simulations with this setup using a rHMC algorithm, which allows us to simulate the model for non-integer $\Nf$.
We are mostly interested in the chiral condensate, which should be non-zero in the broken phase.
Due to the exactly implemented chiral symmetry with the SLAC fermions, it is always zero on every single configuration.
To induce a condensate, we introduced a coupling to a global $U(1)$-symmetric Gross-Neveu term as follows.
We include a new action preserving $U(1)$ symmetry, which reduces to a trivial factor of $1$ in the limit of vanishing coupling. 
With $\Sigma\coloneqq\frac{1}{V} \sum_x \bar\psi_x \psi_x$ and   $\Pi\coloneqq\frac{1}{V} \sum_x \bar\psi_x \gamma_5\psi_x$ it is given by
\begin{equation*}
  1 =\lim_{g_\mathrm{g}\rightarrow 0} \ex{-S_\mathrm{g}} = \lim_{g_\mathrm{g}\rightarrow 0} \ex{\frac{g_\mathrm{g}}{2\Nf} \left(\Sigma^2 - \Pi^2\right)}
\end{equation*}
We perform another Hubbard-Stratonovich transformation, which introduces global fields $\sigma, \pi$ via
\begin{equation*}
  \ex{-S_\mathrm{g}} = \frac{\Nf}{2\pi g_\mathrm{g}} \int_{-\infty}^{\infty}\dx\sigma\int_{-\infty}^{\infty}\dx\pi \, \ex{-\frac{\Nf}{2 g_\mathrm{g}}\left(\sigma^2+\pi^2\right)+\Sigma\sigma+\ii\Pi\pi}.
\end{equation*}
We rescale the fields by $\sqrt{V}$ and define $\lambda_\mathrm{g}\coloneqq\nicefrac{V\Nf}{2g_\mathrm{g}}$.
Now the terms $\Sigma\sigma$ and $\Pi \pi$ can be viewed as part of the Dirac operator.
The resulting full action of the simulated model and its partition sum are
\begin{align*}
  S&=\sum_x \bar\psi\left( \ii \slashed \partial + \slashed V -\frac{1}{\sqrt{V}}\left( \sigma + \ii \gamma_5 \pi\right) \right)\psi+\lambda_\mathrm{Th}\sum_x V_\mu V^\mu + \lambda_\mathrm{g}\left(\sigma^2+\pi^2\right)\coloneqq S\srm{D}+S\srm{Th}+S\srm{g}\\
  Z&=\int \mathcal{D} \bar\psi \mathcal{D}\psi \mathcal{D} V_\mu \dx \sigma \dx \pi\; \ex{-S}\coloneqq \int_0^\infty \dx r\;r \ex{-S\srm{g}(r)} Z(r).
\end{align*}
In the last equation, we defined transformed global fields in a polar coordinate form by $\sigma=r\cos\phi$ and $\pi=r\sin\phi$.
We get a formulation for the absolute value of the chiral condensate by
\begin{equation}
  \chi=\frac{1}{V}\sum_x\left<\bar\psi\ex{\ii\gamma_5\phi}\psi\right>=\int_0^\infty\dx r\;r \ex{-\lambda\srm{g}r^2} \chi_r\label{e:rot_integral}
\end{equation}
Here, $\chi_r$ is defined as
\begin{align}
  \chi_r&=\frac{1}{\sqrt{V}}\frac{\partial \ln Z(r)}{\partial r}= \frac{1}{V Z(r)}\int\Dx\bar\psi\Dx\psi\Dx V_\mu\int_0^{2\pi}\dx\phi\; \sum_x\bar\psi\ex{\ii\gamma_5\phi}\psi \ex{-S\srm{D}-S\srm{Th}}.
  \label{e:rotated_condensate}
\end{align}
In addition, we can define a corresponding susceptibility by inserting $\frac{\partial^2\ln Z(r)}{\partial r^2}$ into \eqref{e:rot_integral}.

\section{Results and Current Work}

\begin{figure}[b]
  \include{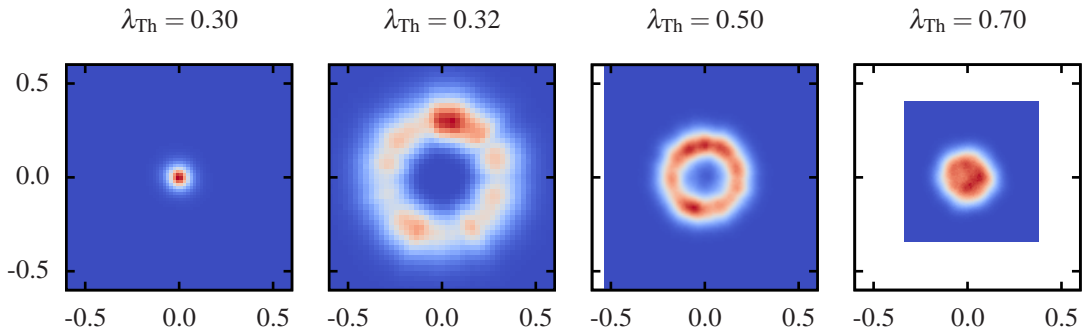}
  \caption{Histograms of the chiral condensate for $\Nf=2$, $\lambda\srm{g}=0.5$ and lattice size 12. The first figure shows the chiral condensate in the unphysical regime, the second and third show the ring shapes in the broken phase and the last is close to the physical phase transition.}
  \label{f:HistLocal}
\end{figure}
\begin{figure}[tbp]
  \begin{minipage}[t]{.49\linewidth}
    \include{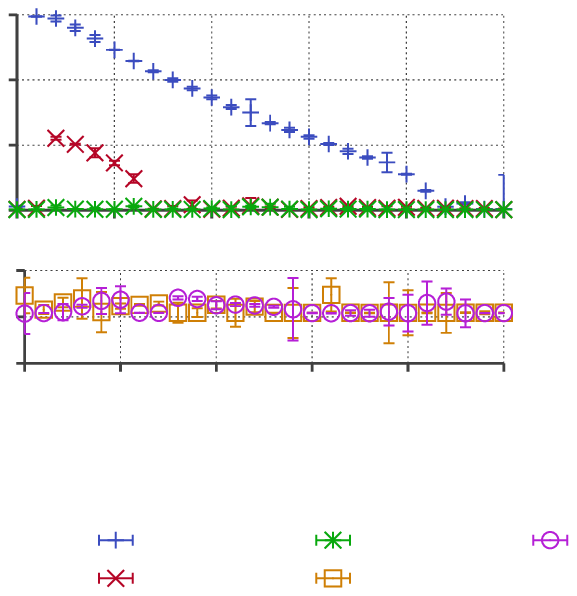}
    \caption{Absolute value of the chiral condensate for $\lambda\srm{g}=0.5$ and size 12. For $\Nf\geq 4$ no non-zero condensate can be observed.}
    \label{f:NfFit}
  \end{minipage}
  \hfill
  \begin{minipage}[t]{.49\linewidth}
    \include{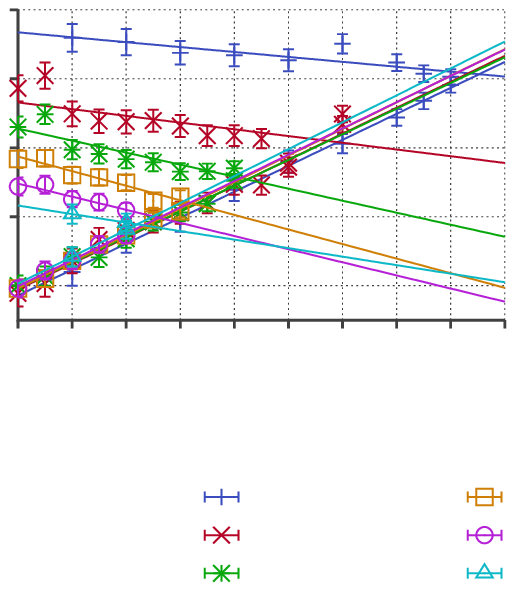}
    \caption{Critical values of the Thirring coupling for physical and artefact transition with linear interpolations.
    For each $\Nf$ the physical coupling lies at larger $\lambda\srm{Th}$ than the artificial one.
    }
    \label{f:NonIntNf}
  \end{minipage}
\end{figure}

The method described in section \ref{s:current_approach} allows us to obtain histograms of real and imaginary parts of the chiral condensate like in figure \ref{f:HistLocal}.
For $\Nf=2$ and $3$ we can see ring shapes in these histograms, indicating a chirally broken phase.
We take the radius of these rings as an estimate for the absolute value of the chiral condensate as in figure \ref{f:NfFit}.
There is a maximal radius for these rings at small inverse couplings with fixed value of $\lambda\srm{Th}\approx 0.17\Nf$.
For increasing $\lambda\srm{Th}$, the size of the rings decreases up to a point, where the rings melt to disks.
This indicates a transition to a chirally symmetric phase.
To the left of the maximal chiral condensate, there is a very sharp decrease in radius and the histograms become very sharply centred points.
Thus, we have a second transition to a phase of zero chiral condensate.
Following the interpretation of \cite{DelDebbio1997-1}, this transition is a sign of the renormalized coupling \eqref{e:renorm_coupling} getting imaginary and thus leading to an unphysical theory.
Therefore, we will call this point near the maximal chiral condensate an artefact transition.
We found evidence in the time series of the chiral condensate, that this transition is of first order.

Simulations on lattices of size\footnote{The lattice size is chosen such that the SLAC derivative gives antiperiodic boundary conditions in the time direction and periodic in the space directions. This requires an even number of points in time direction and an odd number in space directions, see \cite{Wozar2011}. We always took $L\srm{s}=L\srm{t}-1$, such that size 12 refers to a lattice with $12\times11\times11$ points.} 12 and 16 with a fixed global coupling of $\lambda\srm{g}=0.5$ and non-integer values of $\Nf$ between 1.8 and 4 showed, that the two points of phase transitions get closer with increasing $\Nf$.
Figure \ref{f:NonIntNf} shows the positions of both phase transitions depending on $\Nf$ and $\lambda\srm{g}$.
For $\lambda\srm{g}=0.5$ the two critical couplings get very close to each other around $\Nf\approx 3.5$ and we cannot observe any non-vanishing chiral condensate for larger $\Nf$.
Increasing $\lambda\srm{g}$ in order to recover the Thirring model, the point of coincidence of the two phase transitions moves to smaller $\Nf$.
An extrapolation of the intersections of the extrapolating lines in figure \ref{f:NonIntNf} with a fit function $a x^{-b}+c$ yields a limiting value of $\Nfc\mleft(\lambda\srm{g}\rightarrow\infty\mright)=2.05\pm0.05$.

\begin{figure}[tbp]
  \begin{minipage}[t]{.49\textwidth}
    \include{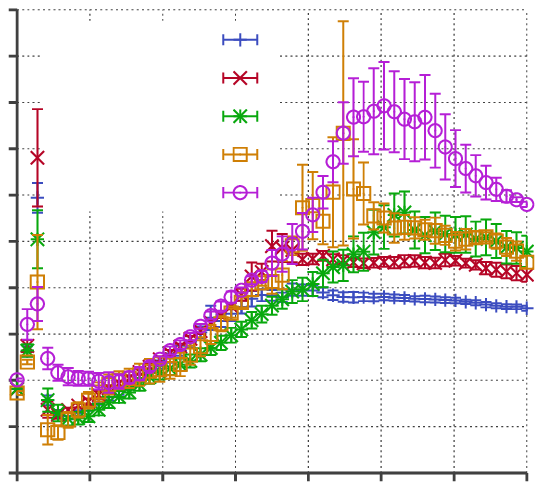}
    \caption{Susceptibility for $\Nf=1$ and $\lambda\srm{g}=0.5$. The sharp peak on the left corresponds to the artefact transition, which is of first order.
    The right peak is at the position of the physical transition.}
    \label{f:SusceptScaling}
  \end{minipage}
  \hfill
  \begin{minipage}[t]{.49\linewidth}
    \include{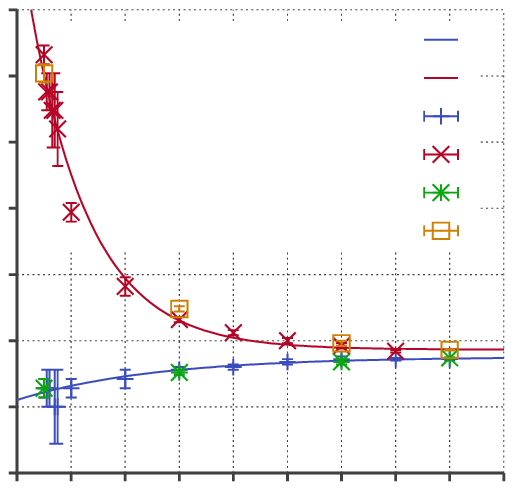}
    \caption{Position of peaks of the susceptibility for $\Nf=1$ with extrapolations in the global coupling for lattice size 12. Note, that there is no second peak for $\lambda\srm{g}=2.0$ at lattice size 12.} 
    \label{f:globalSus}
  \end{minipage}
\end{figure}
Thus, simulations of our coupled model for $\Nf=1$ should show chiral symmetry breaking for all values of the global coupling.
But also in this case, the radius of the histograms gets too small to extract a reliable estimate at $\lambda\srm{g}\approx 1.6$.
Using maxima of the susceptibility (see figure \ref{f:SusceptScaling}) as another indicator of the phase transitions, we were able to determine their positions in $\lambda\srm{Th}$ up to $\lambda\srm{g}=1.8$ as shown in figure \ref{f:globalSus}.
For $\lambda\srm{g}=2.0$ the peak at the position of the physical transition cannot be resolved any more at lattice size 12, but it is visible at size 16.
Thus, to determine if the two phase transitions stay separate as suggested by the extrapolations in figure \ref{f:globalSus}, we have to increase the lattice size, if we want to further increase the global coupling.
At the moment, we cannot draw a reliable conclusion regarding the critical flavour number of the Thirring model.

\section*{Acknowledgements}
D.S. and B.W. were supported by the DFG Research Training Group 1523/2 ``Quantum and Gravitational Fields''. B.W. was supported by the Helmholtz International Center for FAIR within the LOEWE initiative of the State of Hesse.

\printbibliography

\end{document}